\documentclass[prd,aps,twocolumn,floatfix,showpacs,preprintnumbers,amsmath,amssymb]{revtex4}
\usepackage{amsmath}
\usepackage{graphicx}
\usepackage{dcolumn}
\usepackage{bm}
\usepackage{amssymb}
\usepackage{latexsym}

\def\be{\begin{equation}}
\def\ee{\end{equation}}
\def\ba{\begin{eqnarray}}
\def\ea{\end{eqnarray}}

\usepackage{color}

\begin{document}


\title{Cosmological Perturbations in Unimodular Gravity}

\author{Caixia Gao \footnote{Email: cgao1@go.olemiss.edu}}
\affiliation{Department of Physics and Astronomy, University of Mississippi, University, MS 38677, USA}

\author{Robert H. Brandenberger \footnote{Email: rhb@hep.physics.mcgill.ca}}
\affiliation{Department of Physics, McGill University, Montr\'eal, QC H3A 2T8, Canada}

\author{Yifu Cai \footnote{Email: yifucai@physics.mcgill.ca}}
\affiliation{Department of Physics, McGill University, Montr\'eal, QC H3A 2T8, Canada}

\author{Pisin Chen \footnote{Email: chen@slac.stanford.edu}}
\affiliation{Department of Physics \& Graduate Institute of Astrophysics, National Taiwan University, Taipei, Taiwan 10617}
\affiliation{Leung Center for Cosmology and Particle Astrophysics, National Taiwan University, Taipei, Taiwan 10617}
\affiliation{Kavli Institute for Particle Astrophysics and Cosmology, SLAC, \\
Stanford University, Stanford, CA 94305, USA}


\begin{abstract}

We study cosmological perturbation theory within the framework of unimodular gravity. We show that the Lagrangian constraint on the determinant of the metric required by unimodular gravity leads to an extra constraint on the gauge freedom of the metric perturbations. Although the main equation of motion for the gravitational potential remains the same, the shift variable, which is gauge artifact in General Relativity, cannot be set to zero in unimodular gravity. This non-vanishing shift variable affects the propagation of photons throughout the cosmological evolution and therefore modifies the Sachs-Wolfe relation between the relativistic gravitational potential and the microwave temperature anisotropies. However, for adiabatic fluctuations the difference between the result in General Relativity and unimodular gravity is suppressed on large angular scales. Thus, no strong constraints on the theory can be derived.

\end{abstract}

\maketitle

\section{Introduction}

Since the initial discovery by the High-z Supernova Search Team and the Supernova Cosmology Project in 1998 \cite{Riess:1998cb, Perlmutter:1998np}, additional experimental evidence has confirmed that the expansion of our universe is accelerating today. This cosmic acceleration is considered to be one of the most mysterious puzzles in cosmology. The simplest interpretation, in the context of General Relativity (GR), is that the accelerated expansion is induced by a cosmological constant. However, the required value is more than $120$ orders of magnitude smaller than that predicted by quantum field theory with a Planck scale cutoff. Thus, the conflict between data (interpreted as an upper bound on the cosmological constant) and theory already existed before the discovery of accelerated expansion in 1998. It is referred to as the ``cosmological constant problem" and has attracted attention over decades with various solutions, see Refs. \cite{Weinberg:1988cp, Peebles:2002gy, Copeland:2006wr, Frieman:2008sn, Cai:2009zp, Li:2011sd, Chen:2013} for recent reviews.

Unimodular gravity is an alternative formulation of the gravitational theory initiated by Einstein himself nearly one century ago \cite{Einstein:1919}. Its basic idea is that the determinant of the metric is constrained to be a non-dynamical variable. This requirement reduces the symmetry group from invariance under the full group of diffeomorphisms to invariance under the group of unimodular general coordinate transformations (transformations which leave the determinant of the metric unchanged). As a result, the equations of motion governing the dynamics of spacetime are the trace-free Einstein equations, and in this theory the vacuum energy has no direct gravitational effect - vacuum energy does not gravitate, and the cosmological constant is simply an integration constant of the dynamics. Because of this property, it has been suggested that the unimodular gravity theory may relieve the huge discrepancy between the theoretical prediction and the observed value of the cosmological constant in General Relativity \cite{Weinberg:1988cp}. Motivated by this, unimodular gravity has attracted the interest of theorists for years and has been extensively studied in the literature (for example see Refs. \cite{Anderson:1971pn, vanderBij:1981ym, Henneaux:1989zc, Unruh:1988in, Unruh:1989db, Bombelli:1990ke, Ng:1990xz, Finkelstein:2000pg, Alvarez:2005iy, Alvarez:2006uu, Shaposhnikov:2008xb, Smolin:2009ti, Ellis:2010uc, Coley:2011cd, Jain:2011jc, Jain:2012gc, Alvarez:2012px, Eichhorn:2013xr, Ellis:2013uxa, Brown:2013usa, Barcelo:2014mua}).

Provided that the matter fields satisfy the continuity equation, the cosmological background equations of motion derived in unimodular gravity are classically equivalent to those obtained in GR plus an integration constant. Hence, at the background level, cosmology governed by unimodular gravity is degenerate with cosmology in GR in the presence of a cosmological constant. Hence, these two theories are almost indistinguishable in Hubble diagram measurements and in most other dark energy observations.

The question arises, however, as to how unimodular gravity and GR compare when cosmological inhomogeneities are taken into consideration. This is the issue we address in this paper. Most current cosmological observations involve inhomogeneities and anisotropies, and by comparing the evolution of fluctuations in the two theories we can develop tools to distinguish them observationally. Thus, we here develop the theory of cosmological perturbations in unimodular gravity. The theory of cosmological perturbations in GR has been well studied in the literature and is comprehensively reviewed in \cite{Mukhanov:1990me}, and has proven to be very powerful in explaining numerous observational facts concerning Cosmic Microwave Background (CMB) anisotropies \cite{Planck} and Large Scale Structure (LSS) \cite{Percival:2009xn}.

It turns out that, compared with the standard perturbation theory based on GR, that based on unimodular gravity would have one less dynamical equation of motion. In exchange, one obtains an additional constraint equation due to the invariance under unimodular coordinate transformation. It turns out that the main equation of motion for the gauge invariant gravitational potential (the Bardeen variable $\Phi$ \cite{Bardeen}) is the same as in GR, but the gauge freedom of unimodular gravity is less than that in GR. Specifically, in GR there are two scalar gauge modes which can both be set to zero in the conformal Newtonian gauge; however, this gauge choice is no longer possible in unimodular gravity. This is because only one of the two modes can be set to zero, but not both. The second mode should be determined by the constraint equation in terms of the gravitational potential. This non-vanishing mode (we choose the gauge in which it appears as the shift variable) does not change the initial gravitational potential, but it does affect the propagation of photons after decoupling. We study the Sachs-Wolfe (SW) effect \cite{Sachs:1967} and derive the relation between the temperature fluctuations and the gravitational potential, and find that it is modified in unimodular gravity. The non-vanishing shift variable leads to an additional contribution to the temperature anisotropies. However, this extra contribution is suppressed on large angular scales, and hence it does not appear as if interesting constraints on unimodular gravity can be derived.

The paper is organized as follows. In Section II, we briefly review the unimodular gravity theory and list the equations of motion governing the background evolution of the universe. Then in Section III, we study the dynamics of cosmological perturbations in this theory. In particular, we consider matter to be either a single scalar field or a single perfect fluid, and we study the gauge degrees of freedom of the metric perturbations. Afterwards, in Section IV we apply the theory of cosmological perturbations of unimodular gravity to the late time evolution of the universe and study the SW effect that leads to the relation between the temperature fluctuation and the gravitational potential. By comparing with the result in GR, we find an extra term in the SW effect that could be used to distinguish unimodular gravity from GR. In Section V, we summarize our findings and discuss the results.

\section{Brief Review of Unimodular Gravity}

Unimodular gravity is a gravitational theory alternative to Einstein Gravity in which space-time is a Riemannian manifold with a physical measure which is fixed by observations. Specifically, in unimodular gravity the determinant of the metric must obey a specific constraint. The action of unimodular gravity minimally coupled to matter fields is \cite{Finkelstein:2000pg}
\begin{eqnarray}
 S = \int d^4x \sqrt{-g} \left[ \frac{M_p^2}{2}R +\frac{M_p^2}{2}\chi (1-\frac{\xi}{\sqrt{-g}})  +{\cal L}_m \right] ~,
\end{eqnarray}
where $R$ is the Ricci scalar of the physical metric $g_{\mu\nu}$, $M_p$ is the Planck mass, $g$ stands for the determinant of the metric and ${\cal L}_m$ is the Lagrangian of matter fields. $\chi$ is a Lagrangian multiplier and $\xi$ is a fixed function which we take to be a function of time only. One can derive the field equations of the above theory from the action principle. The variation of the action with respect to $\chi$ yields a constraint equation on the metric:
\begin{eqnarray}\label{constraint_g}
 \sqrt{-g} = \xi~.
\end{eqnarray}
Moreover, the variation with respect to the metric gives the Einstein equations
\begin{eqnarray}\label{Einstein_eq}
 R_{\mu\nu} -\frac{R}{2}g_{\mu\nu} -\frac{\chi}{2}g_{\mu\nu} = \frac{1}{M_p^2} T_{\mu\nu}~,
\end{eqnarray}
where we have introduced the stress-energy tensor $T_{\mu\nu}$ as usual. The trace of Eq. \eqref{Einstein_eq} determines the Lagrangian multiplier to be
\begin{eqnarray}\label{chi}
 \chi = -\frac{1}{2} (R+\frac{T}{M_p^2})~,
\end{eqnarray}
where $T\equiv T^\mu_\mu$ is the trace of stress-energy tensor. Substituting \eqref{chi} back into \eqref{Einstein_eq} leads to the modified Einstein equations
\begin{eqnarray}\label{EoM_Einstein}
 \hat{G}_{\mu\nu} = \frac{1}{M_p^2} \hat{T}_{\mu\nu} ~,
\end{eqnarray}
with
\begin{eqnarray}
 \hat{G}_{\mu\nu} \equiv R_{\mu\nu}-\frac{R}{4}g_{\mu\nu} ~,~~~~
 \hat{T}_{\mu\nu} \equiv T_{\mu\nu}-\frac{T}{4}g_{\mu\nu}~.
\end{eqnarray}

Unlike the conventional Einstein Gravity where there are $10$ component equations, in unimodular gravity there are only $9$ independent component equations of motion due to the extra constraint on the determinant of the metric. Thus one can not derive the conservation equation for the stress-energy tensor of matter from the Bianchi identity as can be done in Einstein gravity. {Instead, the conservation equation for the stress-energy tensor
\begin{eqnarray}
 \nabla_\mu T^{\mu\nu} = 0~,
\end{eqnarray}
has to be imposed as an independent assumption. Alternatively, one needs to derive the equations of motion for matter from the variation of the matter field Lagrangian. Recently, unimodular gravity is found can be resulted from nonlinear theory of graviton self-interactions \cite{Barcelo:2014mua}, and the conservation of stress-energy tensor can be derived from the Poincaré invariance.

For illustration, in this work we will consider a universe filled with either a single scalar field or a perfect fluid. First we start with a detailed analysis of perturbations in the case when matter is a single scalar field $\varphi$ with Lagrangian
\begin{eqnarray}
 {\cal L}_m = \frac{1}{2}\nabla_\mu\varphi \nabla^\mu\varphi - V(\varphi)~,
\end{eqnarray}
with $V$ being the field potential. The corresponding equation of motion is the Klein-Gordon equation:
\begin{eqnarray}\label{EoM_KG}
 \nabla_\mu\nabla^\mu \varphi +V_{,\varphi} =0~,
\end{eqnarray}
where $V_{,\varphi}$ denotes the derivative of the scalar field potential with respect to $\varphi$.

To study the cosmological implications, we start with a spatially flat Friedmann-Robertson-Walker (FRW) metric,
\begin{align}
 ds^2 &= dt^2-a^2(t)\gamma_{ij}dx^i dx^j  \nonumber\\
 &= a^2(\tau)(d\tau^2-\gamma_{ij}dx^i dx^j) ~,
\end{align}
where $t$ is cosmic time and $\tau$ is comoving time defined by $d\tau=a^{-1}dt$. Note that  the use of comoving time usually simplifies the equations for cosmological fluctuations and will hence be used in the following sections. For completeness, we
present the background equations of motion in both frames.

In terms of cosmic time, from \eqref{EoM_Einstein} we obtain the single background equation of motion
\begin{eqnarray}\label{eom_dotH}
 \dot{H} = -4\pi G\dot\varphi^2 ~,
\end{eqnarray}
where the dot denotes the derivative with respect to $t$, and $H\equiv \dot{a}/a$ is the Hubble rate characterizing the expanding of the universe. From the above equation, one can read off that the potential energy (and thus the vacuum energy) does not explicitly appear in this background equation. However, when combined with the Klein-Gordon equation \eqref{EoM_KG}, one recovers the standard Friedman equation
\begin{eqnarray}\label{eom_H2}
 H^2 = \frac{8\pi G}{3} (\frac{1}{2}\dot\varphi^2 + V) +\Lambda_{u} ~,
\end{eqnarray}
which allows for a constant of integration $\Lambda_u$ that is purely determined by astronomical observations. Based on the fact that the scalar field potential does not appear in the modified Friedmann equations one can argue that the ``old" cosmological constant problem (using Weinberg's \cite{Weinberg2} terminology), namely the question as to why the huge vacuum energy is invisible in cosmology, is resolved.  The problem is replaced by the presence of a of an arbitrary integration constant whose value has to be fixed by observations.

The same background equation of motion \eqref{eom_dotH} can be reexpressed as follows
\begin{equation}\label{eom_Hprime}
 {\cal H}'-{\cal H}^2=-4\pi G {\varphi '}^2 ~,
\end{equation}
when we transfer to comoving time. Here the prime denotes the time derivative with respect to $\tau$ and ${\cal H}=a'/a$ represents for the comoving Hubble parameter. Additionally, the Klein-Gordon equation is given, in terms of comoving time, by,
\begin{equation}\label{eom_KG_eta}
 \varphi'' +2{\cal H}{\varphi'} +a^2 V_{,\varphi} = 0 ~.
\end{equation}

\section{Cosmological Perturbations}\label{Sec:Pert_primordial}

In this section we develop the theory of linear perturbations in unimodular gravity about the cosmological background. Since we are not breaking spatial rotational symmetry of the background (since $\zeta$ is a function of time only), we can, as usual, decompose the metric perturbations into scalar, vector and tensor fluctuations, which at linear order evolve independently. As we consider matter to be given by a scalar field or a perfect fluid, it is only the metric perturbation of scalar type which at linear order couple directly to matter and hence are of most interest (they will be the dominant fluctuations on large scales in an expanding universe).

Following \cite{Mukhanov:1990me}, one can express the FRW metric including scalar type cosmological perturbations as
\begin{align}\label{scme}
 ds^2 &= a^2(\tau)\,\big[\,(1+2\phi)dt^2-2B_{;i}dx^i d\tau \nonumber\\
 & -((1-2\psi)\gamma_{ij}+2E_{;ij})dx^i dx^j\,\big],
\end{align}
where the subscript ``$;$" denotes the covariant derivative with respect to the $3$-dimensional spatial background metric. There are four metric perturbation variables $\phi$, $\psi$, $B$, $E$ which are functions of both space and time.

\subsection{A universe filled with a single scalar field}

First, we consider the case that the universe consists of a single scalar $\varphi$ which is perturbed by the field fluctuation $\delta\varphi(t, x^i)$. The perturbed Einstein equations of unimodular gravity take the form
\begin{eqnarray}
 \delta \hat{G}^{\mu}_{\nu} = 8\pi G\delta\hat{T}^{\mu}_{\nu} ~.
\end{eqnarray}
Performing a straightforward calculation, one obtains the following non-vanishing component equations
\begin{align}
 \psi'' +{\cal H}(\phi-\psi)' &+\frac{\nabla^2}{3} \big[\phi +2\psi -{\cal H}(B-E') +(B-E')' \big] \nonumber\\
   &=8\pi G \varphi' \delta\varphi' ~, \nonumber \\
 \psi'' +{\cal H}(\phi-\psi)' &+\nabla^2 \big[ \phi +{\cal H}(B-E') +(B-E')' \big] -2D_{,ii} \nonumber\\
   &=8\pi G \varphi' \delta\varphi' ~, \nonumber \\
 (\psi' +{\cal H}\phi)_{,i} &= 4\pi G \varphi' \delta\varphi_{,i} ~, \nonumber \\
 D_{,ij} &=0 ~,
\end{align}
where $D=\phi-\psi+2{\cal H}(B-E')+B'-E''$ and $\nabla^2$ is the Laplace operator on the constant time hyper-surface. The above equations of motion correspond to the $(00)$, $(ii)$, $(0i)$ and $(ij)$ (off diagonal) components of the perturbed Einstein equations, respectively. Again, one finds that the scalar field potential does not show up in the equations of motion. At first glance, it seems like that the potential does not leave any effect on the spacetime geometry. However, this is not true since we also must consider the perturbed Klein-Gordon equation which reads
\begin{align}
 & \delta\varphi'' +2{\cal H}\delta\varphi' -\nabla^2\delta\varphi +a^2 V_{,\varphi\varphi}\delta\varphi -\varphi'(\phi+3\psi)' \nonumber\\
 & ~~~~~~~~~~~~~ +2a^2 V_{,\varphi}\phi -\varphi'\nabla^2(B-E')=0 ~.
\end{align}
As we will show, the potential $V$ affects the dynamics of the metric perturbations due to the combination of the perturbed Klein-Gordon and the Einstein equations.

Different from GR, in unimodular gravity there is an extra constraint on the determinant of the metric, namely $g^{\mu\nu}\delta g_{\mu\nu}=0$, which at the level of linear fluctuations implies,
\begin{equation}\label{constraint_pert}
 \nabla^2 E+\phi-3\psi=0.
\end{equation}
As a result, the gauge freedom is different from that in GR. The preferred gauge choices in GR such as longitudinal gauge and synchronous gauge are not available in unimodular gravity. This can be easily seen, as shown in the following subsection.

\subsection{Gauge freedom}

Since we are only interested in the scalar type metric perturbations, we focus on those diffeomorphisms which preserve the scalar form. The most general ones can be described by two functions $\xi^0$ and $\xi$ which induce the following coordinate transformation,
\begin{align}
 \tau &\rightarrow \tilde{\tau}=\tau+\xi^0(\tau, \vec{x}) ~, \nonumber\\
 x^i &\rightarrow \tilde{x}^i=x^i+\gamma^{ij}\xi_{,j}(\tau, \vec{x}) ~.
\end{align}
In GR, $\xi^0$ and $\xi$ are two independent functions. However, in unimodular gravity they are related by,
\begin{align}
 \nabla^2 \xi+\xi^0{}'+4{\cal H} \xi^0 =0 ~.
\end{align}

The above coordinate transformation induce the following changes in the metric variables,
\begin{align}
 &\tilde{\phi}=\phi-(a'/a)\xi^0-\xi^0{}' ~,~~ \tilde{\psi}=\psi+(a'/a)\xi^0 ~, \nonumber\\
 &\tilde{B}=B+\xi^0-\xi' ~,~~ \tilde{E}=E-\xi ~.
\end{align}
Neither $\delta \hat{G}^\mu_\nu$ or $\delta \hat{T}^\mu_\nu$ is gauge-invariant. However, as in GR we can construct gauge-invariant variables
\begin{align}\label{PhiPsi}
 \Phi &=\phi+(1/a)[(B-E')a]' ~, \nonumber\\
 \Psi &=\psi-(a'/a)(B-E') ~,
\end{align}
and correspondingly we have the gauge-invariant trace-free Einstein tensor $\delta {\hat{G}^{(gi)}}{}^{\mu}_{\nu}$:
\begin{align}
 \delta{\hat{G}^{(gi)}}{}^{0}_{0} &=\delta \hat{G}^0_0+(^{(0)}\hat{G}^0_0)'(B-E') ~, \nonumber\\
 \delta{\hat{G}^{(gi)}}{}^{0}_{i} &=\delta \hat{G}^0_i+(^{(0)}\hat{G}^0_0-\frac{1}{3}{}^{(0)}\hat{G}^k_k)(B-E')_{,i} ~,\nonumber\\
 \delta{\hat{G}^{(gi)}}{}^{i}_{j} &=\delta \hat{G}^i_j+(^{(0)}\hat{G}^i_j)'(B-E') ~.
\end{align}
Similarly, we can obtain the gauge-invariable trace-free stress-energy tensor $\delta{\hat{T}^{(gi)}}{}^{\mu}_{\nu}$. We then obtain the gauge-invariant equations for cosmological perturbations,
\begin{align}
 {\Psi}''+{\cal H}({\Phi}'-{\Psi}')+\frac{1}{3}\nabla^2(\Phi+2\Psi) &=8\pi G{\varphi}'\delta{\varphi}^{(gi)}{}',\nonumber\\
 {\Psi}''+{\cal H}({\Phi}'-{\Psi}')+\nabla^2\Phi-2(\Phi-\Psi)_{,ii} &=8\pi G{\varphi}'\delta{\varphi}^{(gi)}{}',\nonumber\\
 ({\cal H}\Phi+{\Psi}')_{,i} &=4\pi G{\varphi}' \delta \varphi^{(gi)}_{,i},\nonumber\\
 (\Phi-\Psi)_{,ij} &=0,
\end{align}
where $\delta{\varphi}^{(gi)}=\delta \varphi+\varphi '(B-E')$.

From the off diagonal component equation it follows, as in GR, that $\Phi=\Psi$. Then the remaining perturbation equations simplify to:
\begin{align}\label{eom_PhiB}
 &{\Phi}''+\nabla^2 \Phi = 8\pi G{\varphi}'\delta{\varphi}^{(gi)}{}',\nonumber\\
 &{\cal H}\Phi+{\Phi}'=4\pi G{\varphi}' \delta \varphi^{(gi)}, .
\end{align}
In addition, there is the unimodular constraint equation
\begin{eqnarray}\label{constraint_UG}
 (B'-E'') +4{\cal H}(B-E') -\nabla^2 E +2\Phi=0 ~,
\end{eqnarray}
which is derived from the constraint on the determinant \eqref{constraint_pert}. The two equations of \eqref{eom_PhiB} do not depend on the $B$ and $E$ modes, and they can be identified as two of the three total perturbation equations from GR. Although it looks like that GR has one more equation, only two of the three are dynamically independent. Thus, we conclude that the dynamics of the gauge invariant relativistic gravitational potential in unimodular gravity is the same as in GR. We can see this in more detail as follows. Making use of the second equation of \eqref{eom_PhiB}  to express $\varphi^{(gi)}$ in terms of $\Phi'$ and $\Phi$ and then substituting the result into the first equation, one then obtains a second-order differential equation for $\Phi$, which is written as
\begin{equation}
 \Phi'' +2({\cal H}-\frac{\varphi ''}{\varphi '})\Phi' -\nabla^2\Phi +2({\cal H}'-{\cal H}\frac{\varphi ''}{\varphi '}) \Phi =0~.
\end{equation}
This main perturbation equation is exactly the same as the one derived in GR. Consequently, we can follow the standard procedure in traditional cosmological perturbation theory and define the quantity $\zeta$ by
\begin{equation}\label{zeta}
 \zeta=\Phi+\frac{2}{3}\frac{{\dot \Phi}+{H}\Phi}{{H}(1+w)}~,~~ w=\frac{P}{\epsilon} ~,
\end{equation}
with $P$ being the pressure, $\epsilon$ the energy density and $w$ the equation of state, respectively. This quantity is conserved when the wavelength of the mode being considered is larger than the Hubble radius \footnote{This result is true provided that there are no entropy fluctuations and that $w \neq -1$. The result was initially discussed in \cite{Bardeen, BST, BK, Lyth}, and has more recently been extended beyond linear perturbation theory (see e.g. \cite{Langlois}.}. The Klein-Gordon equation for the gauge invariant field fluctuation $\delta \varphi^{(gi)}$ is given by
\begin{align}
 &\delta \varphi^{(gi)}{}''+2{\cal H} \delta \varphi^{(gi)}{}' -\nabla^2\delta \varphi^{(gi)}+V_{,\varphi \varphi}a^2 \delta \varphi^{(gi)}\nonumber\\
 &-4 \varphi '\Phi '+2V_{,\varphi}a^2 \Phi=0 ~.
\end{align}
which also decouples from $B$ and $E$. Thus, this system of perturbation equations is closed and contains only one independent scalar degree of freedom, which can be taken to be either $\Phi$ or $\delta \varphi^{(gi)}$ \footnote{On scales larger than the Hubble radius it is more physical to focus on the relativistic potential $\Phi$.}

However, because in unimodular gravity there is only one scalar gauge degree of freedom, it is not possible to impose either longitudinal gauge or synchronous gauge, two of the gauges most commonly used in cosmological perturbation theory. Because of the reduced gauge symmetry we can define another gauge invariant perturbation variable:
\begin{eqnarray}\label{Sigma}
 \Sigma = B-E'+\int^\tau d\tilde\tau[4{\cal H} (B-E')-\nabla^2E]~,
\end{eqnarray}
which can be set to zero in GR via the gauge transformation which is lost in unimodular gravity. Now, in unimodular gravity we have a new constraint equation which relates the fluctuation variables and yields
\begin{equation}
\Sigma' = -2\Phi \, .
\end{equation}
Thus, $\Sigma$ is not an independent dynamical degree of freedom.

We can use the single scalar gauge degree of freedom to set either $B=0$ or $B=E'$. This choice will simplify the perturbation analysis but does not change the solution of the second order partial differential equation for $\Phi$. It  only affects the relationship between $B$, $E$ and $\Phi$, as shown in the unimodular constraint equation \eqref{constraint_pert} (or \eqref{constraint_UG}). In order to compare with the perturbation analysis in longitudinal gauge in GR, we will from now on specifically choose the $E=0$ gauge in unimodular gravity. Then the constraint equation reduces to
\begin{equation}\label{eom_Bp}
 2\Phi+4 {\cal H}B+B'=0~.
\end{equation}\label{B}
Combining this constraint with the relation
\begin{equation}
 \Phi = \Psi = \psi-{\cal H}B=\phi+{\cal H}B+B' \, ,
\end{equation}
one gets
\begin{equation}
 \phi=3\psi ~,
\end{equation}
in this case.

As in the case of GR, when we consider perturbation modes with wavelengths larger than the Hubble radius during a phase of cosmological evolution with constant background equation of state, then $\Phi$ is almost constant. In this case, the constraint equation can be integrated and yields an expression for $B$ in terms of $\Phi$. Namely, in the radiation-dominated period (when ${\cal H}\simeq 1/\tau$), one obtains the approximate solution in the long wavelength regime as
\begin{eqnarray}\label{Bso_r}
 B \simeq -\frac{2\tau}{5}\Phi \big|_{\rm radiation} ~.
\end{eqnarray}
Moreover, in the matter-dominated period (when ${\cal H} \simeq 2/\tau$), the solution takes the form of
\begin{equation}\label{Bso}
 B \simeq -\frac{2\tau}{9} \Phi \big|_{\rm matter} ~.
\end{equation}
In the above two approximate solutions, we have neglected the decaying modes and the dominant modes are growing linearly in comoving time as the universe expands. This scalar type metric perturbation cannot be removed through a gauge choice in unimodular gravity and thus leads to the possibility of observationally distinguishing unimodular gravity from GR. This will be analyzed in Section IV.

\subsection{A universe filled with a single perfect fluid}

For completeness we consider in this subsection the case when matter consists of a perfect fluid. The stress-energy tensor of a perfect fluid takes the form
\begin{equation}
 T^\alpha_\beta = (\epsilon +P)u^\alpha u_\beta -P \delta^\alpha_\beta ~,
\end{equation}
where the definitions of $\epsilon$ and $P$ are the same as in the previous subsection, and $u^\alpha$ is the four-velocity of the perfect fluid.

Following the analysis in the case of scalar field matter, we see that in unimodular gravity there is only one background equation of motion which in terms of cosmic time is given by,
\begin{eqnarray}
 \dot{H} = -4\pi G(\epsilon+P)~,
\end{eqnarray}
while in terms of comoving time it takes the form
\begin{equation}\label{fe}
 {\cal H}'-{\cal H}^2 = -4 \pi G a^2 (\epsilon +P) ~.
\end{equation}
Notice that in GR the single perfect fluid satisfies the stress-energy conservation as a consequence of the Bianchi identity, but in unimodular gravity it is no longer a consequence of this geometrical identity but must be introduced as a separate assumption \cite{Ellis:2013uxa}. The conservation of stress-energy tensor in the FRW background gives
\begin{equation}\label{fc}
 \epsilon' +3{\cal H}(\epsilon +P)=0 ~.
\end{equation}
Multiplying $6 {\cal H}$ on both sides of \eqref{fe} and then combining with \eqref{fc}, we easily derive
\begin{equation}
 \big( \frac{{\cal H}^2}{a^2} \big)'=\frac{8\pi G}{3} \epsilon' ~,
\end{equation}
from which we again recover the standard Friedman equation plus an arbitrary integration constant. This constant may be fixed by the current measurements of the Hubble diagram if one expects it to be responsible for the late-time acceleration of the universe.

We only consider metric perturbations of scalar type, since they are the only one to couple to matter. The perturbed stress-energy tensor of matter has the form
\begin{equation}
 \delta T^\mu _\nu= \left( \begin{array}{cc}
 \delta \epsilon & -(\epsilon +P)a^{-1}u_i \\
 (\epsilon + P)a u^i &-\delta P \delta_{ij} \end{array} \right) ~,
 \end{equation}
where $\delta \epsilon$ and $\delta P$ are the perturbed energy density and pressure, respectively.

As has been shown in the previous subsection, at the background level the regular Einstein equation can be recovered from the trace-free Einstein equation combined with the continuity equation. This has been verified for a universe filled by a single scalar field, as well as for the perfect fluid of this subsection.

For a perfect fluid, the continuity equation for the fluctuations is valid. Therefore, the field equations for the fluctuations will be the same as those in GR with one extra constraint due to the unimodular coordinate condition. In analogy with the perturbation analysis of the scalar field case, it follows that the equation of motion for the relativistic potential is,
\begin{align}
 \Phi'' +(2{\cal H}'-2{\cal H}^2)\Phi &+\nabla^2\Phi \nonumber\\
  &= 4\pi Ga^2 ( \delta \epsilon^{(gi)} + \delta P^{(gi)}) ~, \nonumber \\
 (a \Phi)'_{,i} &=4\pi Ga^2 (\epsilon +P)\delta u^{(gi)}_i ~,
\end{align}
where
\begin{eqnarray}
 \delta \epsilon^{(gi)} &=&\delta \epsilon +\epsilon'_0(B-E') ~, \nonumber\\
 \delta p^{(gi)} &=& \delta p +p'_0(B-E') ~, \nonumber\\
 \delta u^{(gi)}_i &=& \delta u_i + a(B-E')_{,i} ~.
\end{eqnarray}
In addition, the unimodular constraint equation is given by \eqref{constraint_UG}. Therefore, the gauge-invariant quantity $\zeta$ is conserved on large wavelengths also in the case of the universe filled with a perfect fluid.

\section{Cosmological Perturbations after the Primordial Epoch}\label{Sec:Pert_radiation}

In this section we will discuss the implications of the modified cosmological fluctuation equations for present day cosmological observations. We are specifically interested in the relation between the amplitude of the CMB anisotropies and that of the large-scale structure fluctuations which are given by the relativistic potential. We will find that unimodular gravity predicts a different relationship. Since on large angular scales the GR prediction has now been verified to very good accuracy by current observations, the difference between the predictions of unimodular gravity and GR are tightly constrained. To study this issue, we must generalize the Sachs-Wolfe \cite{Sachs:1967} analysis to the case of unimodular gravity. We find that the difference is suppressed on large angular scales, and hence does not provide an interesting constraint on unimodular gravity.

In the following, we must follow the photon geodesics from the time of decoupling to the present. The initial conditions are determined in the baryon-radiation plasma era before recombination. Hence, we will first start with an analysis of this plasma, followed by a review of the Boltzmann equation which describes the photon gas, before turning to the propagation of photons to the present time.

\subsection{Baryon-radiation plasma}

Before recombination, baryon matter and radiation are strongly coupled and hence the cosmic system can be approximately treated as a single imperfect fluid. The corresponding energy-momentum tensor is given by \cite{Mukhanov:2005sc},
\begin{equation}
 T^\alpha_\beta =(\epsilon +P)u^\alpha u_\beta -P\, \delta^\alpha _{\beta}-\mbox{\boldmath{$\eta$}}({\cal P}^\alpha _\gamma u_\beta {}^{;\gamma}+{\cal P}^\gamma_\beta u^\alpha{}_{;\gamma}-\frac{2}{3}{\cal P}^\alpha_\beta u^\gamma{}_{;\gamma})~,
\end{equation}
where $\mbox{\boldmath{$\eta$}}$ is the shear viscosity coefficient and ${\cal P}^\alpha_\beta=\delta^\alpha _\beta-u^\alpha u_\beta$ is the projection operator. The energy-momentum tensor is assumed to be conserved. Accordingly, the perturbed $(00)$ component equation at the linear order can be calculated as,
\begin{equation}
 \delta \epsilon' + 3{\cal H}(\delta \epsilon +\delta P) -3(\epsilon +P)\psi' +a(\epsilon+P)u^i_{,i}-2\mbox{\boldmath{$\eta$}}{\cal H}\nabla^2 \frac{B}{a}=0 ~.
\end{equation}
Note that if we keep the $B$ term in the metric, then the shear viscosity appears in the perturbed continuity equation. This is required in the theory of unimodular gravity as discussed in the previous section. This term is proportional to $\nabla^2$ and thus is negligible for perturbation modes with long wavelengths, but for modes with short wavelengths it would become significant.

If we consider the long-wavelength regime, when the baryons are non-relativistic, the energy conservation law is valid for both the baryon matter and the radiation separately. In this case, for the perturbation of the radiation sector, we get
\begin{equation}{\label{fradiation}}
 (\delta_\gamma-4 \psi)'+\frac{4}{3}a u^i_{,i}=0 ~,
\end{equation}
where we define the fractional radiation density perturbation as $\delta_\gamma \equiv \frac{\delta \epsilon_\gamma} {\epsilon_\gamma}$. The $(0i)$ component of the energy-momentum tensor is given by
\begin{equation}\label{T^i_0}
 T^i_0 = a(\epsilon +p)u^i -\mbox{\boldmath{$\eta$}}{\cal H} \frac{B_{,i}}{a} ~,
\end{equation}
and again the shear viscosity appears due to the existence of the $B$ term. On large length scales, one can neglect the last term in \eqref{T^i_0} and derive the following useful relation
\begin{equation}\label{tra}
 T^i_{0,i} = \frac{4}{3}\epsilon_\gamma u_0 u^i_{,i}=(4\psi-\delta_\gamma)'\epsilon_\gamma ~,
\end{equation}
where \eqref{fradiation} has been applied.

\subsection{The Boltzmann equation}

During the period of recombination, then as more and more hydrogen atoms form and the temperature drops, photons cease to interact with matter. Afterwards, we can view the photons as a gas of noninteracting identical particles described by kinetic equations. The distribution function $f$, characterizing the number density $\textrm{N}$ of photons in one-particle phase space, is defined by
\begin{equation}
 \textrm{d} N=f(\tau, x^i, p_i)\textrm{d}^3 x \textrm{d}^3 p ~,
\end{equation}
where the position of the particle is given by $x^i(\tau)$ and its 3-momentum is $p_i(\tau)$. For noninteracting particles the distribution function obeys the collisionless Boltzmann equation
\begin{equation}\label{Boltz}
 \frac{D f(\tau, x^i(\tau), p_i(\tau))}{D \tau}\equiv\frac{\partial f}{\partial \tau} +\frac{\textrm{d} x^i}{\textrm{d}\tau} \frac{\partial f}{\partial x^i} +\frac{\textrm{d}p_i}{\textrm{d}\tau} \frac{\partial f}{\partial p_i} =0 ~,
\end{equation}
where $\textrm{d} x^i / \textrm{d} \tau$ and $\textrm{d}p_i/\textrm{d}\tau$ are the derivatives calculated along the geodesics.

For an observer with 4-velocity $u^{\alpha}$ in an arbitrary coordinate system and a photon with 4-momentum $p_\alpha$, the frequency of this photon measured by the observer is given by
\begin{equation}
 \omega = p_\alpha u^\alpha \, .
\end{equation}
If the radiation coming to an observer from different directions
\begin{equation}
 l^i \equiv -\frac{p_i}{{\Sigma p_i^2}^{1/2}} ~,
\end{equation}
has a Planck spectrum, then the distribution function is
\begin{equation}
 f = f(\frac{\omega}{T})\equiv \frac{2}{exp(\omega/T(\tau, x^i, l^i))-1} ~,
\end{equation}
which depends not only on the direction $l^i$ but also on the observer's location $x^i$ at the moment of comoving time $\tau$. In a nearly isotropic universe, the temperature can be decomposed as
\begin{equation}
 T(x^\alpha,l^i) = \bar{T}(\tau)+\delta T(x^\alpha, l^i) ~,
\end{equation}
with $\bar{T}$ being the average background temperature and $\delta T$ the temperature fluctuation.

\subsection{The Sachs-Wolfe effect}

By solving the Boltzmann equation for freely propagating radiation we can find the relationship between the temperature fluctuations and the gravitational potential. We work in the $E=0$ gauge as we have in the previous section. Thus the metric perturbations of scalar type involve only $\phi$, $\psi$ and $B$. We start with the geodesic equations for radiation in an arbitrarily curved spacetime,
\begin{eqnarray}{\label{geodesic}}
 \frac{\textrm{d}x^\alpha}{\textrm{d}\lambda} =p^\alpha ~,~~ \frac{\textrm{d}p_\alpha}{\textrm{d}\lambda} =\frac{1}{2}\frac{\partial g_{\gamma \delta}}{\partial x^\alpha}p^\gamma p^\delta ~,
\end{eqnarray}
where $\lambda$ is an affine parameter along the geodesic. For photons we have $p^\alpha p_\alpha=0$. Using this relation, then, up to the first order in the metric perturbations, one obtains
\begin{align}
 p^0 &= \frac{p}{a^2}(1+\psi-\phi) ~,\nonumber\\
 p_0 &= p ~ (1+\psi+\phi-l^i B_{,i}) ~,
\end{align}
where we have introduced $p \equiv |{\bf p}| = (\Sigma {p_i}^2)^{1/2}$. According to \eqref{geodesic}, we have
\begin{align}\label{gex}
 \frac{\textrm{d}x^i}{\textrm{d}\tau} &= \frac{p^i}{p^0} =l^i ~ (1 +\psi +\phi)- B_{,i} ~,\nonumber\\
 \frac{\textrm{d}p_i}{\textrm{d}\tau} &= \frac{1}{2}\frac{\partial g_{\gamma\delta}}{\partial x^i} \frac{p^\gamma p^\delta}{p^0} = p\,(\psi_{,i}+\phi_{,i} -l^j B_{,ji}) ~.
\end{align}

Therefore, the Boltzmann equation takes the form
\begin{align}
 \frac{\partial f}{\partial \tau} &+ \big[ l^i\,(1+\phi +\psi)- B_{,i} \big] \frac{\partial f}{\partial x^i} \nonumber\\
 &+ p\,(\psi_{,i}+\phi_{,i} -l^j B_{,ji}) \frac{\partial f}{\partial p_i}=0 ~.
\end{align}
As $f$ is a function of only the redshifted frequency
\begin{eqnarray}\label{y_para}
 y\equiv \frac{\omega}{T} =\frac{p_0}{T\sqrt{g_{00}}}
 \simeq \frac{1}{\bar{T} a}[p(1+\psi-\frac{\delta T}{\bar{T}})+B_{,i}p_i] ~,
\end{eqnarray}
then at the background level the Boltzmann equation simply yields $(\bar{T} a)'=0$. Thus, it yields to the well-known relation that in a homogeneous universe the temperature is inversely proportional to the scale factor.

By expanding to linear order in cosmological perturbations, the Boltzmann equation becomes
\begin{equation}\label{Boltz_1st}
 (\frac{\partial}{\partial \tau} +l^i\frac{\partial}{\partial x^i}) (\phi+\frac{\delta T}{T_0}+B^{'})
 =\frac{\partial}{\partial \tau}(2\Phi) ~,
\end{equation}
where we have applied the expressions \eqref{PhiPsi} and the gauge choice $E=0$ in the right hand side of the above equation. This equation determines the dynamics of the temperature fluctuations of the CMB. At the moment of recombination, the universe is already in the matter-dominated phase. Recall that in this phase we have $B'\simeq -2\Phi/9$, which follows from the solution \eqref{Bso} for the perturbation modes on large wavelengths. Since the gravitational potential is constant on super-Hubble scales in the matter-dominated phase, then the right hand side of \eqref{Boltz_1st} vanishes, and hence we derive the generalized SW effect
\begin{equation}\label{SW}
 (\phi+\frac{\delta T}{\bar{T}}+B^{'}) = {\rm const}~,
\end{equation}
along null geodesics. Comparing with the traditional SW effect \cite{Sachs:1967} in GR, we notice that the nonzero metric perturbation $B$ required by unimodular gravity contributes to the CMB anisotropies, and one might expect that this would lead to significant observational effects.

\subsection{Initial conditions}

Consider again the geodesics of the CMB photons arriving from the direction $l^i$ seen by an observer at the present time $\tau_0$ located at $x^i_0$. Making use of the leading terms of Eq. \eqref{gex}, one obtains
\begin{equation}\label{cor}
 x^i(\tau) \simeq x_0^i+l^i(\tau-\tau_0) ~.
\end{equation}
Then the fractional temperature fluctuation $\frac{\delta T}{T}$ in the direction $l^i$ on the sky seen today is related to that emitted from the moment of recombination through the following equation:
\begin{align}\label{tem}
 \frac{\delta T}{T}(\tau_0, x^i_0, l^i) &= \frac{\delta T}{T}(\tau_{rec}, x^i_{rec}, l^i) +\phi(\tau_{rec}, x^i_{rec} ) -\phi(\tau_0, x^i_0) \nonumber\\
 & + B^{'}(\tau_{rec}, x^i_{rec})- B^{'}(\tau_0, x^i_0) ~,
\end{align}
where $\tau_{rec}$ corresponds to the moment of recombination and $x^i_{rec} \equiv x^i(\tau_r)$ can be identified by Eq. \eqref{cor}. Since we are interested in the $l^i$ dependence of the temperature fluctuations and the $\phi(\tau_0, x^i_0)$ term only contributes to the monopole component, we will neglect this term from now on. As a result, the angular dependence of the CMB anisotropies seen today $(\frac{\delta T}{T})_0$ is determined by the following three terms: (1) the initial temperature fluctuations on the last scattering surface; (2) the amplitude of the metric perturbation in the $g_{00}$ component (the redshift term); (3) the amplitude of the metric perturbation in the $g_{0i}$ component along the direction $l^i$ (the Doppler term).

The first contribution, $(\frac{\delta T}{T})_{rec}$, can be evaluated in terms of the metric perturbations and the fluctuations of the radiation energy density $\delta_\gamma = \frac{\delta \epsilon_\gamma}{\epsilon_\gamma}$ on the last scattering surface. To derive the precise relation, one uses the matching conditions between the hydrodynamic energy-momentum tensor, which describes the radiation before decoupling, and the kinetic energy-momentum tensor, which characterizes the gas of free photons after decoupling,
\begin{equation}\label{kin}
 T^\alpha_\beta=\frac{1}{\sqrt {-g}}\int f \frac{p^\alpha p_\beta}{p^0}\textrm{d}^3 p ~.
\end{equation}
Substituting the perturbed metric into the above equation and taking a Planckian distribution, we can easily obtain the $(00)$ component of the kinetic energy-momentum tensor. The result is
\begin{eqnarray}
 T^0_0 \simeq \bar{T}^4 \int \textrm{d}y \textrm{d}^2l ~ [1+4\frac{\delta T}{T} + 3l^iB_{,i}] f(y)y^3 ~,
\end{eqnarray}
with the parameter $y$ being defined in \eqref{y_para}. The integration over $y$ can be done straightforwardly and the result combined with $4\pi \bar{T}^4$ represents the energy density of the photon gas after recombination. This expression ought to match continuously to the $(00)$ component of the hydrodynamic energy-momentum tensor:
\begin{equation}
 T^0_0 = \epsilon_\gamma (1+\delta_\gamma) ~,
\end{equation}
at the moment of recombination. This matching condition implies
\begin{equation}\label{gamma}
 \delta_\gamma=\int (4\frac{\delta T}{T} + 3 l^i B_{,i})\frac{\textrm{d}^2l}{4\pi}
\end{equation}
Similarly, one can derive from \eqref{kin} the $(0i)$ component of the kinetic energy-momentum tensor as
\begin{equation}
 T^i_0 \simeq \epsilon_\gamma \int(4 l^i \frac{\delta T}{T}-2 l^i l^j B_{,j}-B_{,i})\frac{\textrm{d}^2 l}{4\pi}.
\end{equation}
By calculating the divergence of this term and comparing it to the divergence of the hydrodynamical energy-momentum tensor for radiation before recombination as provided in \eqref{tra}, we get following the second matching condition
\begin{equation}\label{gammap}
 \delta'_\gamma \simeq \int \big[ -3 l^i \nabla_i B_{,j}l^j+\nabla_i B_{,i}-4 l^i\nabla_i (\frac{\delta T}{T}) \big] \frac{\textrm{d}^2 l}{4 \pi} ~,
\end{equation}
where we have made use of the constancy of $\phi$ on super-Hubble scales in the matter dominated period, i.e. $\phi'(\tau_{rec})=0$.

It is convenient to study the dynamics of cosmological perturbations in Fourier space since the Fourier modes evolve independently at linear order. In order to satisfy the two above matching relations \eqref{gamma} and \eqref{gammap}, we find that in Fourier space the fractional temperature anisotropies are related to the density fluctuations of radiation and the metric perturbations as follows,
\begin{align}
 (\frac{\delta T}{T})_{\bf k}(\tau_{rec}, l^i) &= \frac{1}{4}\delta_{\bf k}
 +\frac{1}{4}\frac{3\,i}{k^2}(k_m l^m) \delta'_{\bf k} \nonumber\\
 &
 -\frac{3}{4} (l^m k_m) B_{\bf k}(\tau_{rec}, l^i ) ~.
\end{align}
Substituting the above relation into the Fourier form of \eqref{tem}, we can derive the final expression for the temperature fluctuations in the direction $l^i$ as observed at the comoving location ${\bf x}_0$, which turns out to be:
\begin{align}\label{tef}
 & \frac{\delta T}{T}(\tau_0, {\bf x}_0, l^i) = \int \frac{\textrm{d}^3k}{(2 \pi)^{3/2}} \bigg[ \big[ \phi_{\bf k} -\frac{3}{4}({\bf k}\cdot {\bf l})B_{\bf k} + B_{\bf k}^{'} + \frac{\delta_{\bf k}}{4} \big] \nonumber\\
 & ~~
 -\frac{3\,\delta'_{\bf k}}{4 k^2}\frac{\partial}{\partial \tau_0} \bigg]_{\tau_{rec}} e^{i{\bf k}\cdot[{\bf x}_0 + {\bf l}(\tau_{rec} -\tau_0)]}  ~,
\end{align}
where $k \equiv |{\bf k}|$ and ${\bf k}\cdot {\bf l} \equiv k_i l^i$. The terms appearing in the fist line on the right hand side of Eq. \eqref{tef} represent the combined result from the initial inhomogeneities in the radiation energy density itself and the SW effect. The time-derivative term in the second line is related to the velocity of the baryon-radiation plasma at recombination.

\subsection{Anisotropies on large angular scales}

The density fluctuations of radiation are related to the gravitational potential $\psi$ through Eq. \eqref{fradiation}. We neglect the velocity term in \eqref{fradiation}, which is a good approximation in the long-wavelength regime, and then integrate this equation to arrive the following relation:
\begin{equation}\label{cal_C}
 \delta_\gamma - 4 \psi = {\cal C} ~,
\end{equation}
where ${\cal C}$ is an integration constant to be determined. Consider the universe during the epoch of radiation domination. The gravitational potential stays constant on super Hubble scales and hence its value is inherited from the primordial era (such as the one generated during inflation or an early universe period in one of the alternatives to inflation \footnote{See e.g. \cite{RHBrevs} for a recent review of some alternatives to inflation.}). If the fluctuations are purely adiabatic then there is the well known relation
\begin{equation}
 \delta^{(gi)}_\gamma \simeq -2 \Phi (\tau \ll \tau_{eq}) \equiv -2 \Phi_{in} ~,
\end{equation}
where the subscript ``$\textrm{in}$" refers to some moment in the early universe before the time of last scattering. We thus see that the radiation density fluctuation is given by the gravitational potential. If we choose the gauge $E=0$ as we did in Section \ref{Sec:Pert_primordial}, then we can use the relation for the gauge variant potential $\Phi = \psi -{\cal H}B$, together with $ \delta^{(gi)}_\gamma = \delta_\gamma- 4 {\cal H}B$ . Then, we can determine the integration constant ${\cal C}$ with the result
\begin{eqnarray}
 {\cal C} = -6 \Phi_{in} ~. 
\end{eqnarray}
Note that if the $B$ term were already important in either the primordial era or the radiation-dominated phase, the above coefficient would obtain an extra contribution, and correspondingly the modification to the SW effect could be even  more significant.

After the moment of matter-radiation equality, the cold matter becomes dominant and the equation of state of matter changes. The leads to a change in the value of the gravitational potential $\Phi$ on large scales by a factor of $9/10$ (this comes from the conservation of $\zeta$, see \eqref{zeta}). Afterwards $\Phi$ remains constant and so we have,
\begin{equation}
 \Phi (\tau \gg \tau_{eq}) \simeq \frac{9}{10}\Phi_{in} ~.
\end{equation}
Taking into account that during the period of radiation domination there is \eqref{Bso_r},
we find that, when the universe evolves to the recombination, Eq. \eqref{cal_C} yields
\begin{align}
 \delta_\gamma(\tau_{rec}) &= {\cal C}+4 \psi(\tau_{rec}) \nonumber\\
 &\simeq -\frac{8}{3}\psi(\tau_{rec}) +4{\cal H}(\tau_{rec}) B(\tau_{rec}) ~.
\end{align}

Substituting the above expression into Eq. \eqref{tef}, and then applying the approximation that $\delta_{\bf k}' \simeq 0$ on large length scales during the radiation-dominated phase, we then obtain the following formula for the resulting temperature fluctuations
\begin{align}\label{deltaT_T}
 \frac{\delta T}{T}(\tau_0, {\bf x}_0, l^i ) \simeq \frac{1}{3}\Phi(\tau_{rec}, {\bf x}_0-{\bf l}\tau_0 ) +\Delta_{\rm UG} ~,
\end{align}
in terms of the gauge invariant variable $\Phi$. The first term on the right hand side of the above expression  corresponds to the result obtained from in GR which says that the amplitude of temperature fluctuation amplitude is given by one third of the gravitational potential on the last scattering surface from which the photons were produced. The second term is the correction term which appears in unimodular gravity and which is a consequence of the non-vanishing quantity $B$. It takes the form,
\begin{align}\label{rm_UG}
 \Delta_{\rm UG} &= \int \frac{\textrm{d}^3k}{(2 \pi)^{3/2}}
\bigg[ - (\frac{3}{4}({\bf k}\cdot {\bf l}) B_{\bf k}  \bigg]_{\tau_{rec}}
\nonumber\\ &\  \times e^{i{\bf k}\cdot({\bf x}_0+{\bf l}(\tau_{rec} -\tau_0))}
 ~.
\end{align}
We find that the effect of including the shift $B$ on large scales is only dipole. The prediction from unimodular gravity on CMB is then indistinguishable on these scales from that from GR based on current measurements.

We have neglected the contribution of radiation to the gravitational potential at recombination and the integrated SW effect, which may lead to important observational signatures. The amplitude of the matter power spectrum has recently been measured quite accurately, in particular via comparisons of weak gravitational lensing with galaxy clustering (see e.g. \cite{weak}), from SZ cluster studies (see e.g. \cite{Holder}) and from CMB observations (see e.g. \cite{Planck}). Based on current data, the ``bias" parameter $b$ which measures the amplitude of the gravitational potential relative to that of the distribution of visible galaxies, has been determined to an accuracy of better than $3 \%$ (1 sigma confidence) \cite{Planck}. Hence, it is worth while studying fluctuations in unimodular gravity in more detail than what we have done so far.

\section{Conclusion}

The idea of unimodular gravity has recently been discussed extensively in the literature since it resolves the ``old" cosmological constant problem, i.e., the problem of explaining why the large expected value of quantum vacuum energy does not gravitate (a problem that involves quantum physics), by replacing it with a classical problem of fixing an integration constant in the solution of the field equation.
In unimodular gravity, the symmetry group of the theory is reduced from invariance under the full space-time diffeomorphism group to that under unimodular coordinate transformations. The equations of motion for space-time become the trace-free Einstein equations and thus any potential energy, including vacuum energy from the matter fields, is decoupled from the gravitational equations. However, the derivative of a scalar field potential appears in the matter equations and thus can still can affect the evolution of space-time. It turns out that the background equations of motion for cosmology obtained in unimodular gravity are classically the same as those in GR with the addition of an integration constant which can play the role of the cosmological constant. This leads to the interesting question as to whether the theory of unimodular gravity can be distinguished from GR by experiments. There have been attempts at finding a difference by taking into account quantum effects (see e.g. \cite{Unruh:1989db, Alvarez:2005iy, Smolin:2009ti, Eichhorn:2013xr}). So far, however, almost all the results are of theoretical interest only and far from being close to any currently practical experimental test.

In the present paper, we have studied whether unimodular gravity can be distinguished from GR at the level of cosmological inhomogeneities. We have developed the theory of cosmological perturbations for unimodular gravity, with particular emphasis on the gauge freedom of metric perturbations of scalar type under the group of unimodular coordinate transformations. Our results show that the equation of motion for the gravitational potential is unchanged compared with the result in GR. However, there exists another metric perturbation variable which cannot be set to zero in unimodular gravity, unlike the situation in GR. This is a consequence of the reduced gauge symmetry. On the other hand, the new constraint equation relates this variable to the gravitational potential such that it is not an independent dynamical entity. This variable corresponds to the shift in the perturbed metric, and its value grows in an expanding universe.

We have generalized the Sachs-Wolfe (SW) analysis of the relation between gravitational potential and CMB anisotropies to the case of unimodular gravity. Our results show that the extra metric variable leads both to a modification of the gravitational potential contribution to the CMB anisotropies, and also to a change in the geodesics of light between recombination and the present time. Assuming adiabatic fluctuations, we have shown that on large length scales the relationship between the amplitude of the predicted CMB anisotropies and the amplitude of the gravitational potential differs from the result obtained in GR only by a dipole-like term which is suppressed on large scales. This result was derived under the conservative assumption of neglecting any contribution of the shift during the primordial period before recombination. Since the observational bounds on the difference of the predictions from those obtained in GR are tight, it is worth while to revisit our conservative assumptions.

\begin{acknowledgments}

It is a pleasure to thank Luca Bombelli and Gil Holder for useful discussions. The work of RB and CYF is supported in part by an NSERC Discovery grant and by funds from the Canada Research Chair program.  The work of PC is supported in part by the Taiwan National Science Council under Project No. NSC 101-2923-M-002-006-MY3 and 101-2628-M-002-006, by the Taiwan National Center for Theoretical Sciences (NCTS) and by the US Department of Energy under Contract No. DE- AC03-76SF00515. The work of GCX is supported in part by the Summer Research Assistantship from the Graduate School of the University of Mississippi.

\end{acknowledgments}

\end{document}